\documentstyle[12pt,aasms4]{article}

\def\bmvto{\hbox{$(B\!-\!V)_{T}^{o}\ $}}

\begin{document}

\centerline {STATISTICAL CONNECTIONS BETWEEN THE PROPERTIES OF TYPE Ia
SUPERNOVAE}
\centerline {AND THE B--V COLORS OF THEIR PARENT GALAXIES,}
\centerline {AND THE VALUE OF H$_0$}
\bigskip
\author{David Branch, W.~Romanishin, and E.~Baron}
\affil{Department of Physics and Astronomy\\
University of Oklahoma, Norman, OK 73019}
\authoremail{branch,wjr,baron@phyast.nhn.uoknor.edu}

\begin{abstract}
Statistical connections between the properties of Type Ia supernovae
(SNe Ia) and the B--V colors of their parent galaxies are
established. Compared to SNe Ia in blue galaxies [\bmvto$\la$0.75],
SNe Ia in redder galaxies have (1) a wider dispersion in the
blueshifts of their Si II $\lambda$6355 absorption features, ten days
after maximum light; (2) more rapidly declining light curves; and (3)
lower luminosities.  Even when the spectroscopically peculiar, very
subluminous SNe Ia such as SN 1991bg are disregarded, SNe Ia in red
galaxies are less luminous than those in blue galaxies by about 0.3
magnitudes.

When SNe Ia that are thought to have been significantly extinguished
by dust in their parent galaxies are disregarded, those in blue
galaxies have observational absolute--magnitude dispersions of only
$\sigma_{obs}(M_B)$=0.20 and $\sigma_{obs}(M_V)$=0.17, which implies
that their intrinsic absolute--magnitude dispersions are very small.
We use six SNe Ia whose absolute magnitudes have been calibrated by
means of Cepheids, which also indicate that the intrinsic
absolute--magnitude dispersions of SNe Ia in blue galaxies are very
small, to calibrate SNe Ia in blue galaxies and obtain $\rm
H_0=57\pm4\ km\ s^{-1}\ Mpc^{-1}$. This value is in excellent
agreement with that obtained by Saha et al. (1995b), in spite of the
fact that they do not take into account any dependence of SN Ia
absolute magnitude on the nature of the parent galaxy. 

Some implications of the statistical connections between SNe Ia and
the colors of their parent galaxies, for identifying the progenitor
binary systems of SNe Ia and for using high--redshift SNe Ia to
measure q$_0$, are briefly discussed.
\end{abstract}


\section{INTRODUCTION}

Some evidence for statistical connections between the properties of
Type~Ia supernovae (SNe~Ia) and the properties of their parent
galaxies has been published in recent years.  Branch \& van den Bergh
(1993; here BvdB93) found that SNe~Ia that have the lowest blueshifts
of the absorption minimum of the Si~II $\lambda$6355 doublet, ten days
after maximum light, tend to be in galaxies of early type, E to Sa
(see also Filippenko 1989).  Hamuy et al. (1995a; here Ha95a) and
Vaughan, Branch, \& Perlmutter (1995b) find an apparent tendency for
SNe~Ia in spiral, or blue, galaxies to be mildly overluminous with
respect to those in non--spiral, or red, galaxies.  Taken together,
the two lines of evidence are suggestive, but perhaps not conclusive.
For example, Tammann \& Sandage (1995) find no correlation between the
luminosities of SNe Ia and the morphological types of their parent
galaxies.

Established connections between the properties of SNe Ia and of their
parent galaxies could have small effects on the use of SNe Ia to
measure H$_0$, and could be quite important for the more delicate
matters of using SNe Ia to measure q$_0$ (Perlmutter et al. 1995) and
$\Lambda$ (Goobar \& Perlmutter 1995), and galaxy peculiar velocities
(Riess, Press, \& Kirshner 1995b).  They also might furnish important
clues to the identities of the SN Ia progenitor binary systems
(Ruiz--Lapuente, Canal, \& Burkert 1995; Branch et al. 1995a).  The
main purpose of this paper is to show that a convincing case can now
be made for statistical connections between the properties of SNe Ia
and the B--V colors of their parent galaxies.  In addition, we suggest
a refinement to the standard--candle treatment (Sandage et al. 1992,
1994; Branch \& Miller 1993; Schaefer 1994; Saha et al. 1994, 1995a,b;
Tammann \& Sandage 1995) of SNe Ia whose absolute magnitudes have been
calibrated by means of Cepheids in their parent galaxies, to determine
the value of H$_0$.  Some implications for SN Ia progenitors, and for
using high--redshift SNe Ia to measure q$_0$, also are briefly
discussed.

\section{DATA}

The data that are plotted in Figs. 1 to 4 are listed in Table~1, which
includes all SNe Ia for which the B--V color of the parent galaxy and
at least one of the SN Ia parameters plotted in the figures are
available.  For about two thirds of the galaxies in the table,
corrected ``face-on'' total color indices, \bmvto, were taken from the
RC3 Catalogue (de Vaucouleurs et al. 1991).  RC3 colors were not
available for the parent galaxies of the fairly remote SNe Ia reported
by Ha95a.  For these galaxies, approximations to the \bmvto colors on
the RC3 system were obtained as follows: galaxy colors corrected for
Galactic reddening and K terms were taken from Ha95a; galaxy axial
ratios were measured from Digitized Sky Survey images; types were
taken from Ha95a or, for the few galaxies for which Ha95a list
uncertain or unknown types, approximate ``types'' were assigned
according to the colors listed by Ha95a.  Then, from the axial ratios
and types, differential internal extinction corrections were derived
using the precepts of the RC3 Catalogue.

The SN Ia spectral types are on the system of Branch, Fisher, and
Nugent (1993), who subclassified SNe Ia either as normal, or as
peculiar --- like SN 1991bg, like SN 1986G, or like SN 1991T.  SNe Ia
that are assigned a spectral subtype in Table 1 but that were not in
Branch et al. (1993) are SN 1990af (normal; Hamuy et al. 1993), SN
1992K (like SN 1991bg; Hamuy et al. 1994), SN 1992bo (normal; Ha95a),
and SN 1994D (normal; Patat et al. 1995, Meikle et al. 1995).  Values
of the silicon blueshift parameter, V$_{10}$(Si), are from BvdB93
(erroneous entries for SNe 1963J and 1992A have been corrected) except
that values for SNe 1992K, 1992bo, and 1994D have been taken from
Hamuy et al. (1994), Ha95a, and Patat et al. (1995), respectively.

Values of $\Delta m_{15}$, the decline of the blue light curve during
the first 15 days after the time of blue maximum, are from Phillips
(1993; we reject SN 1971I as being too uncertain), Ha95a, and, for SN
1994D, from Richmond et al. (1995).  Absolute magnitudes, M$_B$ and
M$_V$, are from Vaughan et al. (1995b), who apply strict criteria for
inclusion in their sample: the SN Ia apparent magnitudes must be based
on photoelectric or CCD photometry obtained no earlier than 1972,
apparent magnitudes obtained at Asiago Observatory in the 1970s are
excluded, and the estimated uncertainty in the absolute magnitudes
(for a fixed H$_0$) must be no more than 0.5.  For SNe Ia whose
parent--galaxy radial velocities exceed 2500 km sec$^{-1}$, M$_B$ and
M$_V$ are based on H$_0$=85 km s$^{-1}$ Mpc$^{-1}$ (with allowance for
a Local--Group peculiar velocity of 300 km s$^{-1}$ toward the Virgo
cluster), in order to be consistent with the Tully--Fisher (TF) and
surface--brightness--fluctuation (SBF) distances that are used for
galaxies of late and early type, respectively, that have lower radial
velocities.  Vaughan et al. (1995b) corrected M$_B$ and M$_V$ for
foreground extinction (Burstein \& Heiles 1978) but not for extinction
in the parent galaxies.  Here, for four SNe Ia, we correct for the
{\sl total} (foreground plus parent--galaxy) extinction as given by
Hamuy et al. (1995b): E(B--V)=0.10$\pm$0.05 for SN 1981B;
0.65$\pm$0.10 for SN 1986G; 0.37$\pm$0.03 for SN 1989B; and
0.13$\pm$0.05 for SN 1991T.  The absolute magnitudes of SNe 1984A,
1989M, 1990Y, and 1991M, which are suspected of having been
significantly extinguished in their parent galaxies, but for which we
have applied no corrections, are marked with an asterisk in Table~1.

\section{ANALYSIS}

V$_{10}$(Si) is plotted against parent galaxy \bmvto in Fig.~1.  This
figure generally resembles Fig.~2 of BvdB93, in which galaxy
morphological type, rather than \bmvto, was plotted.  BvdB93 found
that the SNe Ia in their sample that had the seven lowest values of
V$_{10}$(Si) were in galaxies of early type (E to Sa).  Here,
similarly, the SNe Ia having the seven lowest values of V$_{10}$(Si)
are in galaxies having \bmvto$\ge$0.81.  This tendency for the lowest
values of V$_{10}$(Si) to be associated with red galaxies clearly is
significant.  BvdB93 also noted that a few of the SNe Ia with the {\sl
highest} V$_{10}$(Si) values were in galaxies of early type.  Here, a
tendency for the highest values to occur in red galaxies is suggested
more strongly.  The four SNe Ia having V$_{10}$(Si) $>$ 12,200 km
sec$^{-1}$ are in galaxies having \bmvto$\ge$0.78.  BvdB93 cited
evidence that two red galaxies hosting SNe Ia with high V$_{10}$(Si)
values --- NGC 4753 hosting SN 1983G and NGC 5493 hosting SN 1990M ---
have undergone fairly recent star formation.  Therefore, it is
possible that SNe Ia that have exceptionally high values of
V$_{10}$(Si), although they are in fairly red galaxies, actually come
from the youngest populations ($\ga 3 \times 10^7$ yr) that can
produce white dwarfs and SNe Ia.\footnote{The referee has asked us
to comment on the curious situation that although V$_{10}$(Si)
correlates with galaxy color, and SN Ia absolute magnitude correlates
with galaxy color (see below), the available data show no clear
correlation between V$_{10}$(Si) and SN Ia absolute magnitude. Apart
from cautioning that although the Si II blueshift is a convenient
observable, its physical interpretation is not necessarily
straightforward (BvdB93), we can offer the following speculation.  The
weakest and dimmest SNe Ia, such as SN 1991bg, do have low values of
V$_{10}$(Si).  The main reasons that there appears to be no clear
correlation between V$_{10}$(Si) and absolute magnitude are that (1)
the overluminous SN 1991T had a fairly low value of V$_{10}$(Si), and
(2) two of the SNe Ia that have the highest values of V$_{10}$(Si),
SNe 1984A and 1983G, are not especially bright.  SN 1991T is a special
case.  Regarding (2), we speculated above that SNe that the highest
values of V$_{10}$(Si), even though they tend to be in red galaxies,
may have come from the youngest populations that can produce SNe Ia.
We also have included SNe 1984A and 1983G among those that may have
been significantly extinguished in their parent galaxies.  Perhaps SNe
Ia with the highest values of V$_{10}$(Si) {\sl are} intrinsically
bright, but being from young populations and associated with
interstellar matter, they tend to be extinguished by dust in their
parent galaxies.}  In any case, the data establish that SNe Ia in
galaxies redder than \bmvto$\simeq$0.75 show a wider dispersion in
their V$_{10}$(Si) values than do those in bluer galaxies.  The null
hypothesis that the two distributions have the same variance (F--test;
c.f. Sachs 1982) has a probability of 0.006.  A Kolmogorov--Smirnov
test (e.g. Sachs 1982) suggests that the two distributions are drawn
from different populations.  Both statistical tests indicate that the
correct place to divide the populations is indeed as suggested by the
eye, near \bmvto$\approx 0.75$.

The $\Delta m_{15}$ index is plotted against galaxy \bmvto in Fig.~2.
(We are grateful to Bob Tripp, of LBL, for suggesting that we look at
this.)  The trend is obvious: SNe Ia in blue galaxies, \bmvto$\la0.75$,
tend to have slower light curve decline rates than do SNe Ia in redder
galaxies.\footnote{Filippenko (1989) drew attention to the possibility
of photometric differences between SNe Ia in elliptical and spiral
galaxies; his suggestion, based on the data that was available at that
time, was not that SNe Ia in ellipticals decline at a faster rate than
do those in spirals, but that those in ellipticals have a smaller
dispersion in their decline rates than do those in spirals.}  The KS
probability that the two distributions are drawn from the same
population is $10^{-4}$.

M$_B$ and M$_V$ are plotted against galaxy \bmvto in Figs.~3 and 4,
 which suggest that SNe Ia in red galaxies tend to be less luminous
 than SNe Ia in blue galaxies, in agreement with Ha95a.  When all of
 the plotted data are considered, the K--S statistic peaks when the
 sample is divided at \bmvto=0.65 in Fig.~3, and at \bmvto=0.75 in
 Fig.~4.  Even when the SNe Ia that are known to have been
 spectroscopically peculiar are disregarded, the absolute magnitudes
 of SNe Ia in blue and red galaxies appear to be different.  In this
 case the K--S statistic peaks when the sample is divided at
 \bmvto=0.65, in both Figs.~3 and 4, but data are sparse in the region
 $0.65\la$\bmvto$\la 0.75$.  The K--S probability that the two
 distributions are drawn from the same sample is $9\times 10^{-4}$ in
 M$_B$ and 0.02 in M$_V$.\footnote {Some might be tempted to fit
 linear regressions to the data in Fig. 1 to 4, but it seems
 inappropriate to do so because SNe Ia in blue and red galaxies appear
 to come from separate populations.  In any case, since the data are
 scattered the linear fits would be poor.}  In galaxies having
 \bmvto$\le$0.72, when the spectroscopically peculiar SN 1991T and the
 SNe Ia that are suspected of having been strongly extinguished in
 their parent galaxies are disregarded, the remaining 11 SNe Ia have a
 mean $\rm M_B =-18.69 (\pm0.06) +5 log(H_0/85)$, with an {\sl
 observational} dispersion of only $\rm \sigma_{obs}(M_B)=0.20$.  In
 galaxies having \bmvto$\ge$0.87, when the SNe Ia that are known to
 have been spectroscopically peculiar are disregarded, the remaining
 eight SNe Ia have a mean $\rm M_B = -18.36 (\pm 0.08) +5 log
 (H_0/85)$, with $\rm \sigma_{obs} (M_V)=0.23$.  The corresponding
 results for M$_V$ are --18.72 $(\pm 0.05)$, $\rm
 \sigma_{obs}(M_V)=0.16$ for 11 SNe Ia in blue galaxies and $\rm
 -18.44(\pm0.09), \sigma_{obs}(M_V)=0.26$ for eight SNe Ia in red
 galaxies.  (If the sample is divided at \bmvto=0.65, thus shifting
 SNe 1991T and 1990S from blue to red, the corresponding values are
 $\rm M_B=-18.75\pm0.05$ in blue galaxies and $\rm M_B =
 -18.38\pm0.07$ in red galaxies, $\rm M_V = -18.76\pm 0.05$ in blue
 galaxies and $\rm M_V = -18.46\pm 0.07$ in red galaxies.) The
 differences $\Delta M_B =0.3 \pm 0.1$ and $\Delta M_V=0.3\pm0.1$
 between normal SNe Ia in blue and red galaxies appear to be
 significant.\footnote{The brightest SN Ia in a red galaxy is an
 interesting case.  SN 1994D, on the system of Branch et al. (1993),
 was a spectroscopically normal SN Ia, but it was too bright for its
 light curve decline rate (Richmond et al. 1995; Patat et al. 1995).
 Considering that the distance--independent U--B color of SN 1994D was
 anomalously negative (Branch et al. 1995b), the apparent
 overluminosity of SN 1994D for its decline rate and its
 parent--galaxy color probably is real rather than due to an
 erroneously long SBF distance, relative to the distances used for the
 other SNe Ia (cf. Tammann \& Sandage 1995).}  We believe the reason
 that Tammann \& Sandage (1995) did not see a difference between the
 absolute magnitudes of SNe Ia in galaxies of early and late type was
 that they used less restrictive criteria to define their
 absolute--magnitude samples.  Their samples of 41 SNe Ia in B and 37
 in V included, for example, SNe 1970J and 1972J, in elliptical
 galaxies, both of which were measured at Asiago observatory and are
 now known to have been measured too bright (Tsvetkov 1986; Patat
 1995).

We conclude, on the basis of the combined evidence of Figs.~1--4, that
real statistical connections between the properties of SNe Ia and the
colors of their parent galaxies have been established.

\section{H$_0$ FROM CEPHEID--CALIBRATED SNe Ia}

When determining the value of H$_0$ from SNe Ia, the difference
between the absolute magnitudes of SNe Ia in blue and red galaxies is,
at least to first order, automatically taken into account if some SN
Ia distance-independent observable, such as $\Delta m_{15}$ (Ha95a),
light--curve--shape (Riess, Press, \& Kirshner 1995a), or V$_R$(Ca)
(Fisher et al. 1995), is used to standardize the SN Ia absolute
magnitudes.  Here, instead of a SN Ia observable, we use a
parent--galaxy observable, \bmvto.  Since SNe Ia whose absolute
magnitudes are calibrated by Cepheids are, naturally, in blue
galaxies, and since we found above that the absolute magnitude
dispersion among SNe Ia in blue galaxies appears to be very small, we
simply use SNe Ia in blue galaxies as nearly standard candles.  This
circumvents the disagreements (Ha95a; Tammann \& Sandage 1995; Pierce
\& Jacoby 1995; Schaefer 1994, 1995b) concerning the reliability of
the light--curve decline rates of some of the Cepheid--calibrated SNe
Ia.

Table 2 lists absolute magnitudes for six Cepheid--calibrated SNe Ia,
all of which were in blue galaxies.  The absolute magnitudes of SNe
1895B, 1937C, and 1972E are from Table 7 of Saha et al. (1995a), with
a slight adjustment for SN 1895B that reflects our adoption of the
peak B magnitude, 8.26$\pm$0.11, recommended by Schaefer (1995a).
Pierce \& Jacoby (1995) advocate revisions to the peak apparent
magnitudes of SN 1937C that were determined by Schaefer (1994) and
adopted by Saha et al. (1995a), but see Schaefer (1995b).  The
absolute magnitudes of SNe 1981B are based on B=12.04$\pm$0.04 and
V=11.96$\pm$0.04 from Schaefer (1995c), E(B--V)=0.10$\pm$0.05 from
Ha95a, and a true distance modulus of 31.10$\pm$0.20 from Saha et
al. (1995b). The absolute magnitudes of SN 1960F are based on
B=11.77$\pm$0.11 and V=11.51$\pm$0.18 from Schaefer (1995d), the
assumption of negligible extinction (see Schaefer 1995d), and a
(preliminary) distance modulus of 31.15$\pm$0.20 (Saha 1995). The
absolute magnitudes of SN 1989B are based on B=12.34$\pm$0.05 and
V=12.02$\pm$0.05 from Wells et al. (1994), E(B--V)=0.37$\pm$0.03 from
Wells et al. (1994) and Hamuy et al. (1995b), and the recently
obtained Cepheid--based distance of 30.32$\pm$0.16 to M96 (NGC 3368)
in Leo (Tanvir et al. 1995). According to Tully (1987), M96 and NGC
3627, the parent galaxy of SN 1989B, both belong to the Leo Spur,
which happens to ``stretch out close to the plane of the sky" and has
``relatively little depth from our viewing position".  If Tully is
correct, then by measuring the Cepheid distance to M96, Tanvir et
al. have provided a Cepheid calibration of the absolute magnitudes of
SN 1989B.  In the error budget, we have allowed for a difference of
$\pm0.2$ magnitudes in the distance moduli of NGC 3627 and M96, i.e.,
for NGC 3627 we have used a distance modulus of 30.32$\pm$0.26.

The weighted mean M$_B$ for the six SNe Ia of Table 2 is --19.59 $\pm$
0.07, with an observational dispersion about the mean of $\rm
\sigma_{obs}(M_B)=0.17$; the weighted mean M$_V$ for five SNe Ia in
Table 2 is --19.58 $\pm$ 0.04, with $\rm \sigma_{obs}(M_V)=0.09$.
Thus the Cepheid calibrations indicate, in agreement with the
conclusion obtained in \S~3, that the intrinsic dispersions of the
absolute magnitudes of SNe Ia in blue galaxies are very small.
Combining these mean absolute magnitudes with the H$_0$--dependent
absolute magnitudes of SNe Ia in galaxies having \bmvto$\le0.72$ that
were obtained in \S 3, and allowing for an uncertainty of 0.15
magnitudes in the Cepheid zero point, leads to $\rm H_0=56\pm4$ from
M$_B$ and $\rm H_0=57\pm4$ from M$_V$.  (If we used galaxies with
\bmvto$\le 0.65$ instead of 0.75 then the corresponding results would
be H$_0=58\pm4$ from M$_B$ and $\rm H_0=58\pm 4$ from M$_V$.)  These
values are in excellent agreement with those obtained by Saha et
al. (1995b), in spite of the fact that they do not take into account
any dependence of SN Ia absolute magnitude on the nature of the
parent galaxy.  These values also are in very good agreement with
those obtained on the basis of Cepheid--independent, direct physical
calibrations of SNe Ia (e.g., Arnett, Branch, \& Wheeler 1985; Branch
1992; M\"uller \& H\"oflich 1994; Fisher et al. 1995; Nugent et
al. 1995a,b,c; H\"oflich \& Khokhlov 1995; van den Bergh 1995; Branch
et al. 1995b).

\section {SOME IMPLICATIONS FOR PROGENITORS}

The statistical connections between the properties of SNe Ia and of
their parent galaxies are likely to be caused by a dependence of the
properties of SNe Ia on the ages (rather than, for instance, the
metallicities) of their binary progenitor systems (Branch \& van den
Bergh 1993; Ruiz--Lapuente, Canal, \& Burkert 1995).  In view of the
fact that the B--V color of the integrated light of a galaxy can
hardly be a precise indicator of the age of the progenitor system of a
particular SN Ia, it seems possible that SN Ia properties and
progenitor age are strongly related.  As Ha95a point out, colors at
the sites of SNe Ia would be preferable to the integrated colors of
the whole parent galaxies.  Most progenitors of SNe Ia last long
enough that even colors at the SN Ia sites would not necessarily be
the same as the colors at the places where the progenitors were
formed.

How could progenitor age affect the SN Ia properties?  For each of the
candidate progenitor binary systems for SNe Ia (Branch et al. 1995a),
the realization frequency is a particular function of age.  Recent
attempts to calculate such functions, for some of the candidate
systems, have been made, for example, by Tutukov \& Yungelson (1994),
Ruiz--Lapuente et al. (1995), Yungelson et al. (1995), and Canal,
Ruiz--Lapuente, \& Burkert (1995).  It may be that the kind of system,
or the mixture of systems, that produces SNe Ia in young populations
(from about $3 \times 10^7$ to, say, $10^9$ years) differs from the
kind of system, or the mixture of systems, that produces SNe Ia in
older populations.  Or, it could be that even for a single system, the
SN Ia properties depend on age, through, e.g., the cooling time, the
mass accretion rate, the initial carbon to oxygen ratio, or the
rotation, of the white dwarf.  Thus, while it may not be surprising to
see a connection between SN Ia properties and progenitor age, it is
too early to decide just how the effect is produced.

\section {SOME IMPLICATIONS FOR USING HIGH--REDSHIFT SNe Ia TO
MEASURE~q$_0$}

SNe Ia are now being discovered at high redshifts, $0.3\la z \la 0.5$,
(Norgaard--Nielsen et al. 1989; Perlmutter et al. 1994, 1995; Schmidt
et al. 1995), and attempts to predict SN Ia rates as a function of
redshift, for some of the candidate progenitor systems, are beginning
to be attempted (Ruiz--Lapuente et al. 1995; Canal et al. 1995).  To
look back to z$\simeq$0.5 is, for H$_0$=50 and q$_0$=0.5, to look back
about $6 \times 10^9$ yr.  So, if the oldest of the nearby SNe Ia come
from systems whose ages are within $6 \times 10^9$ yr of the age of
the universe, such SNe Ia would not be represented at z$\simeq$0.5.
On the other hand, even the oldest of the nearby SNe Ia may come from
systems that have ages as short as, say, $4 \times 10^9$ yr, in which
case the full range of the nearby SNe Ia may be represented in
high--redshift samples.  In any case, given the tendency for young SNe
Ia to be more luminous than older ones, there may be a tendency for
the high-redshift SNe Ia to be somewhat more luminous that the nearby
ones --- unless the search procedures for high--redshift SNe Ia are
biased in favor of galaxies that have old populations, e.g.,
elliptical galaxies, in which case the opposite might be expected.
Consequently, to measure q$_0$, it will be important to use
distance--independent observables to accurately match the
high--redshift SNe Ia to the nearer ones. A light--curve--shape
observable may not be the best choice for some of the high--redshift
SNe Ia, since it would require accurate photometry at times other than
near the light--curve peak.  Some of the maximum--light photometric SN
Ia observables, such as the (rest--frame) B--V (Vaughan et
al. 1995a,b) or U--B (Branch et al. 1995b), or the maximum--light
spectroscopic SN Ia observables (Nugent et al. 1995c) should prove to
be useful.  The results of this paper suggest that parent--galaxy
observables also may be useful for using SNe Ia to measure q$_0$ and
galaxy peculiar velocities.

\acknowledgements
We are grateful to Abi Saha and Allan Sandage, acting for the HST
Cepheid--SN Ia consortium, for providing a preprint and other data in
advance of publication, to Ramon Canal for pointing out an error in an
earlier version of the manuscript, and to Willy Benz, Mario Hamuy,
Alexei Khokhlov, Saul Perlmutter, Brad Schaefer, Bob Tripp, and
referee Sidney van den Bergh for helpful comments and information.
This work has been supported by NSF grants AST 9417102 and 9417242.

\vfill\eject

\section*{FIGURE CAPTIONS}

\bigskip

FIG.~1.---The Si II absorption blueshift parameter V$_{10}$(Si) is
plotted against parent--galaxy \bmvto.  Filled circles: SNe Ia that are known
to have been spectroscopically normal; circles with crosses: SNe Ia
that are known to have been spectroscopically peculiar; open circles:
SNe Ia for which neither spectroscopic normalcy nor peculiarity has
been established.

\bigskip

FIG.~2.---The light--curve decline parameter $\Delta m_{15}$ is
plotted against parent--galaxy \bmvto.  Symbols are as in Fig.~1.

\bigskip

FIG.~3.---SN Ia absolute magnitude M$_B$ is plotted against
parent--galaxy \bmvto.  Symbols are as in Figs. 1 and 2, except that
vertical arrows denote SNe Ia that are thought to have been strongly
extinguished by dust in their parent galaxies but for which no
correction has been applied.

FIG.~4.---Like Fig.~3, but for M$_V$.


\begin{references}

\reference{} Arnett, W. D., Branch, D., \& Wheeler, J. C. 1985, Nature, 314, 337

\reference{} Branch, D. 1992, ApJ, 392, 35

\reference{} Branch, D., Nugent, P., Baron, E., \& Fisher, A. 1995b,
in ``Thermonuclear Supernovae'', ed. R. Canal, P. Ruiz--Lapuente, \&
J. Isern (Dordrecht: Kluwer), in press

\reference{} Branch, D., Fisher, A., \& Nugent, P. 1993, AJ, 106, 2383

\reference{} Branch, D., Livio, M., Yungelson, L. R., Boffi, F., \&
Baron, E. 1995a, PASP, 107, 1

\reference{} Branch, D. \& Miller, D. L. 1993, ApJ, 405, L5

\reference{} Branch, D., \& van den Bergh, S. 1993, AJ, 105, 2231 (BvdB93)

\reference{} Burstein, D., \& Heiles, C. 1978, ApJ, 225, 40

\reference{} Canal, R., Ruiz--Lapuente, P., \& Burkert, A. 1995, ApJ,
in press

\reference{} de Vaucouleurs, G., de Vaucouleurs, A., Corwin, H. G., Buta,
R. J., Paturel, G., \& Fouqu\'e, P. 1991, Third Reference Catalog of
Bright Galaxies (New York: Springer) (RC3)

\reference{} Filippenko, A. V. 1989, PASP, 101, 588

\reference{} Fisher, A., Branch, D., H\"oflich, P. A., \& Khokhlov, A. 1995,
ApJ 447, L73

\reference{} Goobar, A., \& Perlmutter, S. 1995, ApJ, 450, 14

\reference{} Hamuy, M. et al., 1993, AJ, 106, 2392

\reference{} Hamuy, M., et al., 1994, AJ, 108, 2226

\reference{} Hamuy, M. et al. 1995a, AJ, 109, 1 (Ha95a)

\reference{} Hamuy, M., Phillips, M. M., Maza, J., Suntzeff, N. B.,
Schommer, R. A., \& Avil\'es, R. 1995b, in preparation

\reference{} H\"oflich, P. A., \& Khokhlov, A. 1995, ApJ, in press

\reference{} Meikle, W. P. S., et al. 1995, preprint

\reference{} M\"uller, E., \& H\"oflich, P. 1994, A\&A, 281 51

\reference{} Norgaard--Nielsen, H. U., Hansen, L., Jorgensen, H. E.,
Salamanca, A. A., Ellis, R. S., \& Couch, W. J. 1989, Nature, 339, 523

\reference{} Nugent, P., Baron, E., Hauschildt, P. H., \& Branch,
D. 1995a, ApJ 441, L33

\reference{} Nugent, P., Branch, D., Baron, E., Fisher, A., Vaughan, T.,
\& Hauschildt, P. H. 1995b, Phys. Rev. Lett., 75, 394; erratum: 75, 1874

\reference{} Nugent, P., Phillips, M. M., Baron, E., Branch, D., \&
Hauschildt, P. 1995c, ApJ, in press

\reference{} Patat, F. 1995, personal communication

\reference{} Patat, F. et al., 1995, MNRAS, in press

\reference{} Perlmutter, S. et al., 1994, IAU Circ., No. 5956; erratum:
No. 5958

\reference{} Perlmutter, S. et al., 1995, ApJ, 440, L41

\reference{} Phillips, M. M. 1993, ApJ, 413, L105

\reference{} Pierce, M. J., \& Jacoby, G. H. 1995, AJ, in press

\reference{} Richmond, M., et al. 1995, AJ, 109, 2121

\reference{} Riess, A. G., Press, W. H., \& Kirshner, R. P. 1995a, ApJ,
438, L17

\reference{} Riess, A. G., Press, W. H., \& Kirshner, R. P. 1995b, ApJ,
445, L91

\reference{} Ruiz--Lapuente, P., Burkert, A., \& Canal, R. 1995, ApJ,
447, L69

\reference{} Sachs, L. 1982, Applied Statistics: A Handbook of
Techniques, (New York: Springer-Verlag)

\reference{} Saha, A. 1995, personal communication

\reference{} Saha, A., Labhardt, L., Schwengeler, H.,
Macchetto, F. D., Panagia, N., Sandage, A., \& Tammann, G. A.  1994,
ApJ, 425, 14

\reference{} Saha, A., Sandage, A., Labhardt, L., Schwengeler, H.,
Tammann, G. A., Panagia, N., \& Macchetto, F. D. 1995a, ApJ, 438, 8

\reference{} Saha, A., Sandage, A., Labhardt, L., Tammann, G. A.,
Macchetto, F. D., \& Panagia, N. 1995b, preprint

\reference{} Sandage, A., Saha, A., Tammann, G. A., Panagia, N., \&
Macchetto, F. D. 1992, ApJ, 401, L7

\reference{} Sandage, A., Saha, A., Tammann, G. A., Labhardt, L.,
Schwengeler, H., Panagia, N., \& Macchetto, F. D. 1994, ApJ, 423, L13 

\reference{} Schaefer, B. E. 1994, ApJ, 426, 493

\reference{} Schaefer, B. E. 1995a, ApJ, 447, L13

\reference{} Schaefer, B. E. 1995b, preprint

\reference{} Schaefer, B. E. 1995c, preprint

\reference{} Schaefer, B. E. 1995d, preprint
 
\reference{} Schmidt, B. et al., 1995, IAU Circ., No. 6160

\reference{} Tammann, G. A., \& Sandage, A. 1995, ApJ, 452, 16

\reference{} Tanvir, N. R., Shanks, T., Ferguson, H. C., \& Robinson,
D. R. T. 1995, Nature, 377, 27

\reference{} Tripp, R. D. 1995, personal communication

\reference{} Tsvetkov, D. Yu. 1986, P. Zvezdy, 23, 216

\reference{} Tully, R. B. 1987, ApJ, 321, 280

\reference{} Tutukov, A. V., \& Yungelson, L. R. 1994, MNRAS, 268, 871

\reference{} van den Bergh, S. 1995, ApJ, in press

\reference{} Vaughan, T. E., Branch, D., Miller, D. L., \& Perlmutter,
S. 1995a, ApJ, 439, 558

\reference{} Vaughan, T. E., Branch, D., \& Perlmutter, S. 1995b, in preparation

\reference{} Wells, L. A. et al. 1994, AJ, 108, 2233

\reference{} Yungelson, L. R., Livio, M., Tutukov, A. V., \& Kenyon,
S. J. 1995, ApJ, 447, 656

\end{references}
\end{document}